\documentclass[journal]{IEEEtran}
\usepackage{helvet}
\usepackage{courier}
\usepackage{amsmath}
\usepackage{amssymb}
\usepackage{graphicx}
\usepackage{subfigure}
\usepackage{algorithm}
\usepackage{algorithmic}
\usepackage{multirow}
\usepackage{makecell} 
\usepackage{diagbox} 
\usepackage{cases}

\newtheorem{myPro}{Problem}
\newtheorem{myDef}{Definition}

\begin{document}
%
\title{PCNN: Deep Convolutional Networks for Short-term Traffic Congestion Prediction}
%
%
%

\author{Meng~Chen,Xiaohui~Yu,~\IEEEmembership{Member,~IEEE,}
        and~Yang~Liu,~\IEEEmembership{Member,~IEEE}}

\maketitle

\begin{abstract}
Traffic problems have seriously affected people's life quality and urban development, and forecasting the short-term traffic congestion is of great importance to both individuals and governments. However, understanding and modeling the traffic conditions can be extremely difficult, and our observations from real traffic data reveal that (1) similar traffic congestion patterns exist in the neighboring time slots and on consecutive workdays; (2) the levels of traffic congestion have clear multiscale properties. To capture these characteristics, we propose a novel method named \textbf{PCNN} based on deep \textbf{C}onvolutional \textbf{N}eural \textbf{N}etwork, modeling \textbf{P}eriodic traffic data for short-term traffic congestion prediction. PCNN has two pivotal procedures: time series folding and multi-grained learning. It first temporally folds the time series and constructs a two-dimensional matrix as the network input, such that both the real-time traffic conditions and past traffic patterns are well considered; then with a series of convolutions over the input matrix, it is able to model the local temporal dependency and multiscale traffic patterns. In particular, the global trend of congestion can be addressed at the macroscale; whereas more details and variations of the congestion can be captured at the microscale. Experimental results on a real-world urban traffic dataset confirm that folding time series data into a two-dimensional matrix is effective and PCNN outperforms the baselines significantly for the task of short-term congestion prediction.
\end{abstract}

\begin{IEEEkeywords}
Traffic Congestion Prediction, Periodic Traffic Data, Convolutional Neural Network.
\end{IEEEkeywords}

\section{Introduction}
People are getting increasingly concerned  about traffic congestion, which has seriously affected their life quality and urban development. To monitor real-time traffic conditions, many cities around the world have deployed embedding sensors, e.g., inductive-loop detectors and video image processors, in road networks \cite{chen2015mining}, and GPS (Global Position System)-based services such as Google Maps have been developed to show traffic conditions and even details regarding individual vehicles \cite{zhan2017citywide}. The increasing availability of data from such devices and services has created unique opportunities to predict traffic conditions (e.g., predicting travel speed and traffic volume \cite{abadi2015traffic,xu2014accurate,silva2015predicting}, predicting city-scale traffic flow \cite{zhan2017citywide,zhang2016deep}), benefiting the decision making of individuals and governments. For example, people can adjust their driving routes dynamically and authorities can optimize traffic signal time according to predicted traffic conditions.

Most existing work on predicting traffic conditions has focused on predicting future traffic flows at a given location \cite{habtemichael2016short,lv2015plane} or the travel time on a given road segment \cite{yildirimoglu2013experienced}. In this paper, we target instead at directly forecasting short-term traffic congestion levels for road segments in urban road networks. The reason is that very often people would just like to know how ``jammed" the traffic is going to be in the next minutes or hours on the road segments of their interest, rather than the actual traffic flow values or travel time. In this paper, we define the \textit{congestion level} $c$ for a road segment during a given time slot as $c=\max[0,(t-\overline{t})/\overline{t}]$, where $t$ is the average travel time of vehicles for that segment in that time slot, and $\overline{t}$ is the baseline travel time for the same road segment in ideal traffic conditions. Congestion level is an intuitive way to depict the traffic condition and is suitable for visualization, as it resembles an evaluative meter that is understandable to most people.

\begin{figure}[!htbp]
\centering
\subfigure[Traffic congestion levels of two road segments during one week.]{
\includegraphics[width=0.45\textwidth]{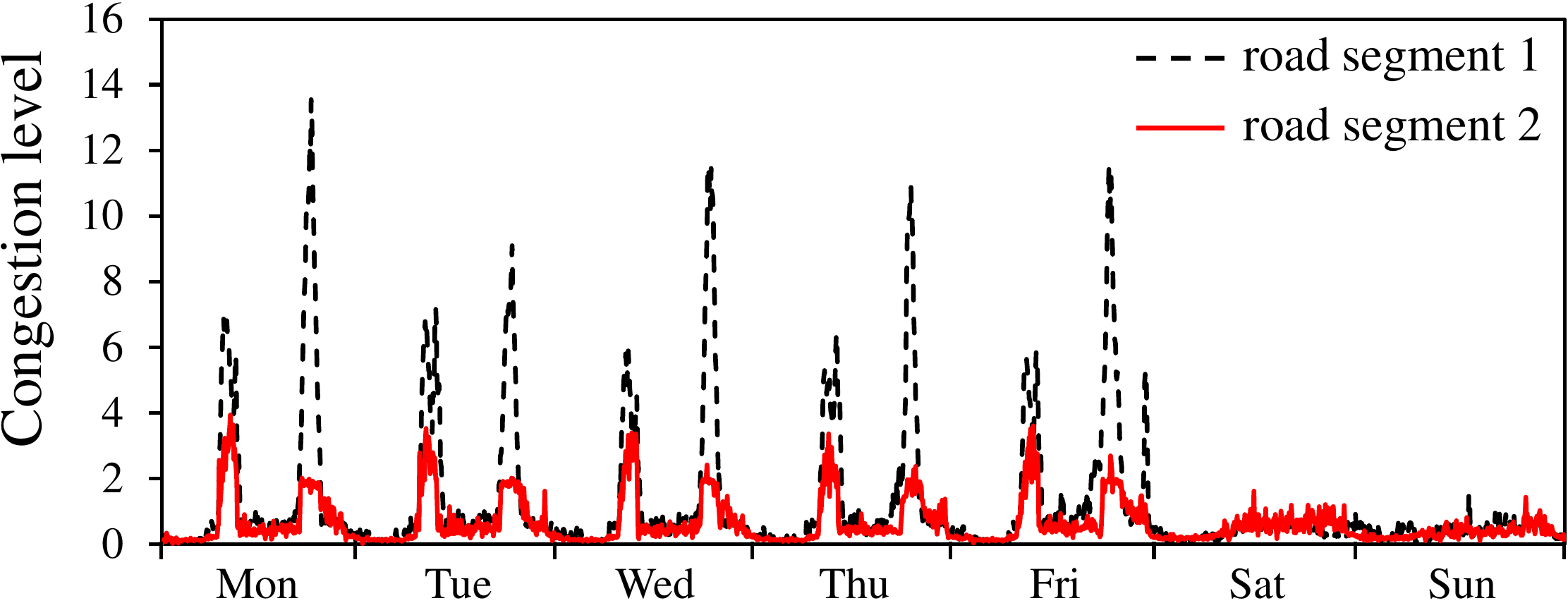}}
\subfigure[Traffic congestion levels of road segment 1 from 6:30 am to 8:00 am.]{
\includegraphics[width=0.45\textwidth]{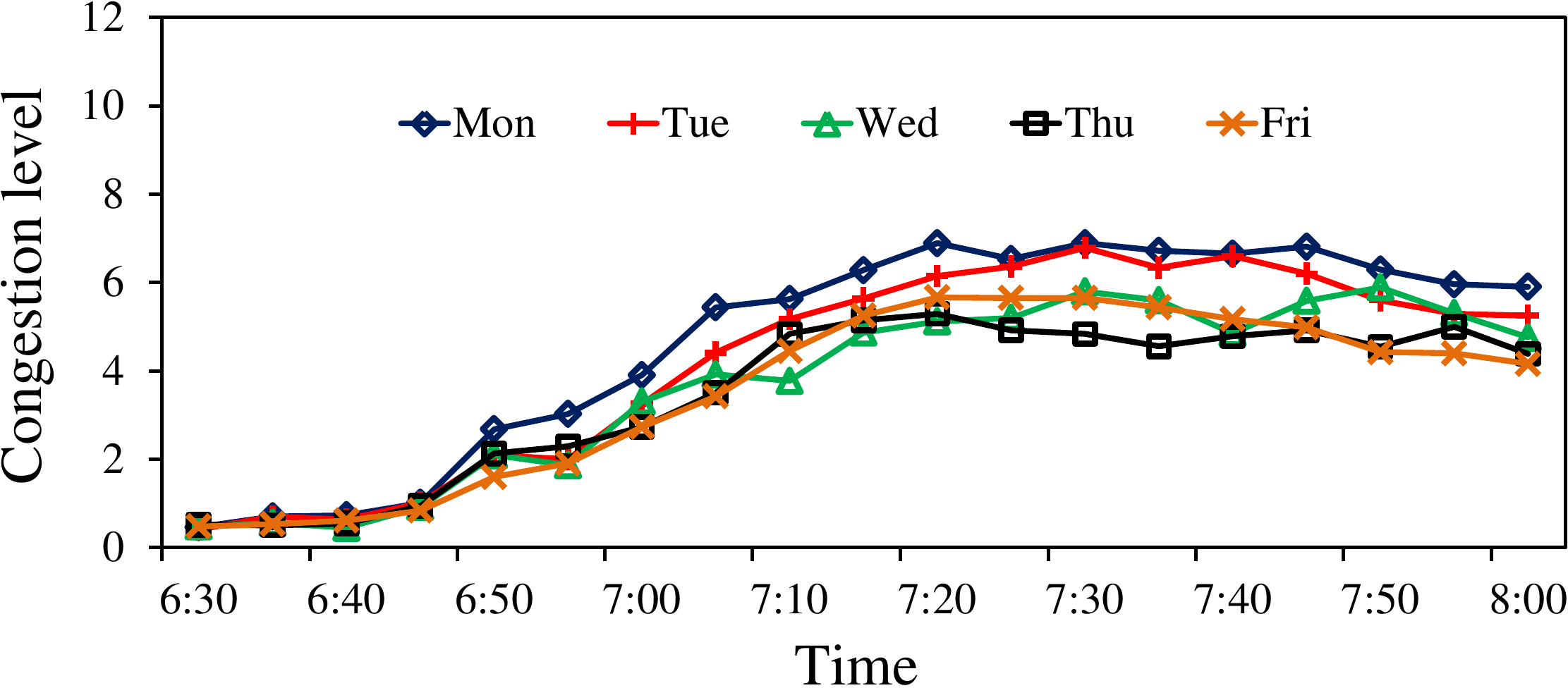}}
\caption{Examples of traffic congestion levels of two road segments.}
\label{oneweek}
\end{figure}

Forecasting traffic congestion levels, however, is filled with challenges, because of a series of complex factors. To demonstrate this, we carefully depict the congestion time series with an urban traffic dataset. We randomly choose two road segments (1 and 2) in Jinan and show their traffic congestion levels during one week in Fig.~\ref{oneweek}(a), and further plot the congestion levels of road segment 1 from 6:30 am to 8:00 am on workdays in Fig.~\ref{oneweek}(b).

\begin{itemize}
\item \textbf{Local coherence}. The traffic congestion level in a time slot has a strong correlation with those in the neighboring time slots, and the correlation diminishes as the temporal distance increases. For example, the traffic conditions of 6 pm may be affected by the congestion occurring at 5 pm, but can be considered free from the influence of the traffic at 8 am of the same day.

\item \textbf{Periodicity}. Traffic congestion levels on different workdays exhibit a temporal periodicity, i.e., repeating a similar pattern roughly every 24 hours. For example, as shown in Fig.~\ref{oneweek}, traffic congestion levels during the same time slots (e.g., morning rush hours) are similar on consecutive workdays, but are different from those in other time periods, e.g., from 11 am to 1 pm, of the same day. Further, our analysis of the real traffic data during a span of six weeks reveals that the congestion levels of a given workday are more similar to those of adjacent days, rather than the same days in other weeks.

\item \textbf{Multiscale property}. Traffic congestion levels have clear multiscale properties. At the microscale, the variation of congestion levels can be observed with precise details, while it is hard to discover the global trend of large temporal scope. In contrast, at the macroscale, the global trend of congestion levels can be easily revealed, while many details are lost. Thus, the traffic congestion level in a given time slot is the result of both global and local effects, and a combination of global trend and local fluctuation may help make better prediction.
\end{itemize}

The properties of local coherence and periodicity imply that the traffic congestion level in a time slot is related to those in the neighboring slots of both the same day and previous days. However, most existing methods \cite{lv2015traffic,lv2015plane} predict future traffic congestion $c_{m,n}$ in the time slot $n$ of day $m$, by taking just the immediately preceding $t$ values and/or the values of the same slot in previous $d$ days into consideration. They fail to consider the similar patterns on preceding workdays, which may degrade the prediction accuracy. In addition, to the best of our knowledge, none of these existing methods consider the multiscale property in making short-term traffic prediction.


\textbf{Present work.} To capture the similar traffic patterns and multiscale congestion properties, we propose \textbf{PCNN}, a \textbf{C}onvolution-based deep \textbf{N}eural \textbf{N}etwork modeling \textbf{P}eriodic traffic data, which converts the one-dimensional data into an image-like input matrix and applies a series of convolutions on it. Specifically, PCNN has two pivotal procedures: time series folding and multi-grained learning. To predict future traffic congestion level $c_{m,n}$, we fold the time series of congestion levels based on the period (i.e., 24 hours), and combine the $2t$ values around the current time slot $n$ in the previous $d$ days with the immediately preceding $t$ values (replicating once) to generate the input matrix with size $(d+1) \times 2t$. Consequently, the two-dimensional matrix contains both the traffic conditions in the immediate past and a large volume of similar historical patterns; thus both local coherence and periodicity are taken into consideration.

Another key contribution of PCNN is in learning a set of multi-grained features. By performing an array of convolutions over the input matrix, PCNN could capture the local temporal dependency and numerous higher-level features. Then, these features are transmitted to the output layer to predict the future traffic congestion levels. Finally, the objective can be efficiently optimized with stochastic gradient descent (SGD) akin to back propagation on the deep convolutional networks.

Our experiments focus on short-term traffic congestion prediction with the real vehicle passage records data in Jinan, China. We contrast the performance of PCNN with state-of-the-art traffic forecasting methods, including regressive models (e.g., ARIMA (autoregressive integrated moving average) \cite{ahmed1979analysis}), pattern recognition methods (e.g., K-NN (K-nearest neighbors) \cite{habtemichael2016short}), and neural networks (e.g., MLP (multilayer perceptrons ) \cite{dougherty1997short} and LSTM (long short-term memory) \cite{tian2015predicting}). Experiments show that PCNN has smaller forecast errors. In addition, we also apply the two-dimensional input matrix to some baselines, and the results demonstrate that the methods with the two-dimensional input performs better than those with original one-dimensional input.

The main contributions can be summarized as follows:
\begin{itemize}
\item Different from existing methods that model traffic patterns with one-dimensional time-series, we propose to fold the traffic data based on the period and model them as a two-dimensional matrix, which considers the traffic conditions in the immediate past and similar historical patterns simultaneously.
\item We propose to apply a series of convolutions on the two-dimensional input matrix to model local temporal dependencies and multi-grained features. We are thus able to estimate the approximate range of future congestion levels at different scales. To the best of our knowledge, this is the first time to apply convolutions on the periodic traffic data.
\item We conduct extensive experiments with the real vehicle passage records in an urban road network to investigate the effectiveness of the proposed PCNN. For the task of congestion prediction, the results demonstrate that the methods with two-dimensional input perform better, and PCNN shows remarkable improvement compared with several baselines.
\end{itemize}

The rest of this paper is organized as follows. Section~\ref{relatedwork} reviews the studies on short-term traffic prediction. Section~\ref{pre} introduces the definition of congestion level and the problem solved in this paper. Section~\ref{pcnn} presents our deep convolutional networks for short-term traffic congestion level prediction.  The experimental results are discussed in Section~\ref{experiment}. Section~\ref{conclusion} concludes this paper.

\section{Related Work}\label{relatedwork}
Traffic congestion prediction can be considered as an extension of short-term traffic forecasting, which is a pivotal application in intelligent transportation systems. See Bolshinsky and Freidman \cite{bolshinsky2012traffic} for a thorough survey on different techniques (e.g., time series models, Markov chain models, non-parametric methods) used for traffic forecasting. Here we only focus on summarizing existing works that are directly related to our study. These works fall into three broad categories, i.e., regressive models, pattern recognition methods and neural networks (NN).

\subsection{Regressive models}
Regressive models are a type of general methods for forecasting time series data. As most traffic data tend to be closely related to their previous values, a special group of regressive models named autoregressive integrated moving average (ARIMA) are usually adopted for short-term traffic prediction. ARIMA is parameterized by three non-negative integers, commonly represented as ARIMA$(p; d; n)$, where $p$ is the number of autoregressive terms, $d$ is the number of nonseasonal differences, and $n$ is the number of lagged forecast errors in the prediction equation. Ahmed and Cook first introduce the model to predict the freeway traffic volume and occupancy time series \cite{ahmed1979analysis}. After that, numerous ARIMA-based variants have been proposed in traffic time series prediction, e.g., seasonal ARIMA \cite{williams1998urban} and space-time ARIMA \cite{kamarianakis2003forecasting}. Chung and Rosalion \cite{chung2001short} systemically compare ARIMA and its variants  with some alternative solutions including regression, historical average, etc. Their results reveal that the above strategies perform reasonably well under normal conditions, but less satisfactory when external changes (e.g., weather, special events) happen.

\subsection{Pattern recognition methods}
Pattern recognition methods have also been applied to short-term traffic forecast, e.g., support vector machines (SVM) \cite{wang2013short}, and K-nearest neighbors \cite{xia2016distributed,habtemichael2016short,yu2016k}. Wang and Shi \cite{wang2013short} integrate Wavelet-Chaos Analysis and SVM regression theory, and construct a new kernel function to capture the non-stationary characteristics of the short-term traffic speed data for prediction. Considering the time-varying and continuous characteristic of traffic flow, Yu et al. \cite{yu2016k} propose a multi-time-step prediction model based on the K-nearest neighbors (K-NN) algorithm for short-term traffic condition prediction. Further, Habtemichael and Cetin \cite{habtemichael2016short} present an enhanced K-NN method using weighted Euclidean distance to identify similar traffic patterns for short-term traffic forecast. In addition, Xia et al. \cite{xia2016distributed} propose a K-NN model in a general MapReduce framework on a Hadoop platform to enhance the efficiency of short-term traffic flow forecasting. However, the pattern recognition methods cannot work well when the number of historical data exhibiting similar patterns is limited, for example, the time slots with extreme traffic congestion are rare, and these methods fail to identify similar patterns for prediction in this case.

\subsection{Neural networks}
Most early studies along this line exploit feed-forward multilayer perceptrons (MLP) \cite{dougherty1997short,clark1993use}, in which the temporal relationships are augmented in the input data during pre-processing. Besides, there are approaches \cite{lingras2001time,abdulhai2002short} adopting dynamic neural networks, e.g., Lingras and Mountford \cite{lingras2001time} use a genetic algorithm to optimize the connections between inputs and hidden layers for traffic volume estimation.

Deep neural network (DNN), which refers to a feed forward neural network with more than one hidden layers, has recently revolutionized the machine learning society, and achieved great success in natural language processing, computer vision, etc.  Convolutional neural network (CNN) and recurrent neural network (RNN) are two main types of DNN architectures. In general, CNN is hierarchical, and originally applied to capture spatial features in image classification \cite{krizhevsky2012imagenet}. RNN exhibits a sequential architecture, and is intuitively plausible for sequence modeling tasks, e.g., language modeling \cite{sutskever2014sequence}. In practice, both kinds of neural networks have been explored to capture spatial and temporal dependencies, and sometimes are even applied simultaneously \cite{xingjian2015convolutional}. Despite the great success of DNN, only few efforts have been made to use it for traffic forecasting. As a representative piece of work, Tian and Pan \cite{tian2015predicting} adopt the LSTM (long short-term memory method) which could determine the optimal length of the input historical data dynamically to make short-term traffic flow prediction.

In this paper, we work on the specific problem of short-term traffic congestion prediction, and introduce a novel convolutional neural network to model the intricate natures of temporal features, including periodicity, local coherence, etc., which have been rarely considered in previous studies. In the same vein, some recent studies attempt to perform the city-scale crowd flow prediction with DNN \cite{li2015traffic,zhang2016deep}, whose objective is to estimate the total traffic of crowds entering/leaving a region during a given time interval. Nonetheless, in contrast to our work that aims to explain the temporal dependencies, they tend to focus on capturing spatial dependencies, as the inflow of one region is affected by outflows of nearby regions in their applications.

\section{Preliminaries}\label{pre}

We first introduce the definition of congestion level and then formally define the problem to be addressed in this paper.

\begin{myDef}[Congestion Level] Given a road segment, we define the \textbf{congestion level} $c_{i,j}$ based on the average travel time $t_{i,j}$ for that segment in time slot $j$ of day $i$ and the baseline travel time $\overline{t}$ for the same road segment, formally as $c_{i,j} = \max(0, (t_{i,j}-\overline{t})/\overline{t})$, where $\overline{t}$ can be estimated using data from periods of light traffic (e.g., after midnight).
\end{myDef}

\begin{myPro}[Short-term Traffic Congestion Forecasting]
For a specific road segment, given a sequence of observed congestion data $\{ c_{i,j} \}$, $i = 0, 1, \ldots, m$, $j = 0, 1, \ldots, n-1$, where $i$ represents the index of day and $j$ is the index of time slot in day $i$, the problem is to predict anticipated traffic congestion level $c_{m,n}$ in the next time slot of day $m$.
\end{myPro}

\begin{figure*}[!ht]
\centering
\includegraphics[width=0.85\textwidth]{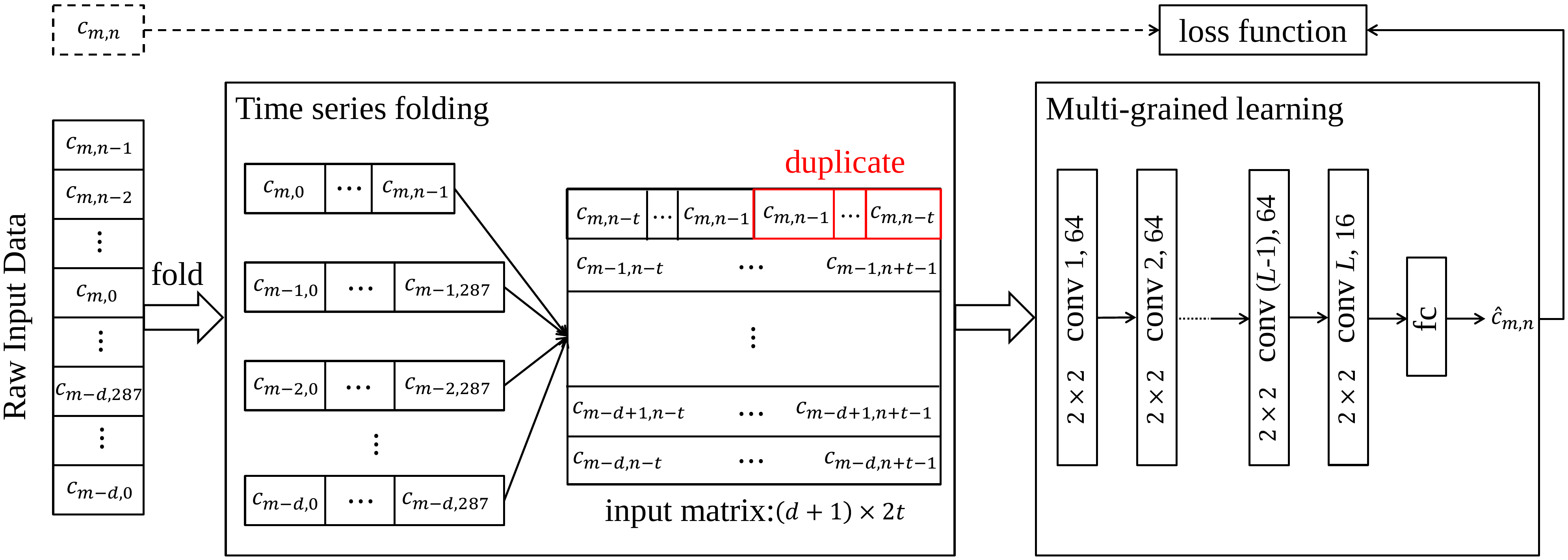}
\caption{Architecture of PCNN. conv: convolutional layer, fc: fully-connected.}
\label{cnn}
\end{figure*}

\section{Deep Convolutional Networks}\label{pcnn}
In this section, we first elaborate the architecture of the proposed neural network, and then provide details of each component. Finally, we introduce the objective function and the method for parameter learning.

\subsection{Overview}
Intuitively, due to local coherence and periodicity, the variations of congestion levels in the neighboring time slots on consecutive workdays are similar. Besides, traffic congestion levels have shown distinct multiscale features. These properties can be effectively handled by the convolutions that have proved effective in capturing the local structural information and multiscale features from pixel-level raw images \cite{krizhevsky2012imagenet}. Inspired by this idea, we propose a tailored method named \textbf{PCNN} to capture the periodic traffic congestion patterns at different scales for predicting short-term traffic congestion levels. Fig.~\ref{cnn} presents the architecture of PCNN, which consists of two major components: time series folding and multi-grained learning. As illustrated in the left part of Fig.~\ref{cnn}, for the raw input data, we take 5 minutes as the size of time slots as an example, and obtain 288 slots in one day. We first choose the historical data from the previous $d$ days and fold them into a two-dimensional matrix. The input matrix is then fed into multi-grained learning component, capturing multiscale features by a series of convolutions. Finally, the output of the final convolutional layer is transmitted to the output layer, yielding the predicted value.

\subsection{Time series folding}

We have observed local coherence and periodicity, but there is no observed evidence that the congestion levels of a particular workday (say, Tuesday) bears higher similarity with those of the same days in previous weeks (past Tuesdays), according to our traffic data. Thus, we only consider the historical values in the preceding $d$ days and fold them into $d$ vectors. Furthermore, the traffic congestion level in a specific time slot has a strong correlation with those in the neighboring time periods. Therefore, to predict the traffic congestion level $c_{m,n}$, where $m$ and $n$ are the indexes of day and time slot, we take the $2t$ values $(\langle c_{m-i,n-t}, \ldots, c_{m-i,n+t-1} \rangle, i \in \{ 1, 2, \ldots, d\})$ around the current slot $n$ in every day into consideration. Note that, it is superior to only using the $d$ values of slot $n$ in previous $d$ days, as the $2t$ time slots exhibit similar traffic congestion patterns in the $d$ days because of local coherence and periodicity. Further, the traffic conditions in the immediate past are pretty important, and we duplicate the congestion levels of the recent $t$ slot to get a vector $\langle c_{m,n-t}, \ldots, c_{m,n-1}, c_{m,n-1}, \ldots, c_{m,n-t} \rangle$ of length $2t$. Afterwards, these vectors are integrated to yield the  $(d+1) \times 2t$ input matrix $\mathbf{X}_{m,n}$:
\begin{equation}
\label{matrix}
\mathbf{X}_{m,n} =
\left[
\begin{matrix}
 c_{m,n-t}    & \cdots  &  c_{m,n-t}  \\
 c_{m-1,n-t}  & \cdots  &  c_{m-1,n+t-1}    \\
 \vdots       & \vdots  &  \vdots \\
 c_{m-d,n-t}  & \cdots  &  c_{m-d,n+t-1}      \\
\end{matrix}
\right].
\end{equation}
In our study, $d$ and $t$ are data-independent, and we will evaluate their effects in the experiments. Further, we use $c_{i,0} (m-d \leq i \leq m)$ to pad those elements of each row vector whose index is less than 0 under condition of $n < t$; similarly, we use $c_{0,j} (n-t \leq j \leq n+t-1)$ to pad each column vector under condition of $m < d$.

\subsection{Multi-grained learning}
In order to capture the multiscale congestion patterns, we decide to apply a series of convolutions on the input matrix. As one convolution only accounts for near dependencies, limited by the size of their kernels, we need to use multiple convolutional layers to model the dependency over a greater time range. Here we do not use pooling operations, but only convolutions, following the suggestion in \cite{zhang2016deep,jain2007supervised}.

Given the input matrix $\mathbf{X}_{m,n}$, we apply the convolutional operation (i.e., conv 1 in Fig.~\ref{cnn}) on it:
\begin{equation}
\mathbf{H}^{(1)} = f ( \mathbf{W}^{(1)} \ast \mathbf{X}_{m,n} + b^{(1)}),
\end{equation}
where $\ast$ denotes the convolution, and $f$ is an activation function, e.g., the rectified linear function (ReLU) \cite{krizhevsky2012imagenet}. $\mathbf{W}^{(1)}$ and $b^{(1)} \in \mathbb{R}$ are the learnable parameters in the first convolutional layer. We then feed $\mathbf{H}^{(1)}$ to the next layer, until the $L$-th layer:
\begin{equation}
\mathbf{H}^{(l)} = f ( \mathbf{W}^{(l)} \ast \mathbf{H}^{l-1} + b^{(l)}), l=2, \ldots, L.
\end{equation}

For each one in the first $(L-1)$ layers, we use 64 filter maps of size $2 \times 2$ at a stride of 1 over the input data. Note that each filter map is replicated across the entire input matrix, and a unit in the filter map has 4 inputs connected to a $2 \times 2$ area in the input matrix, called the receptive field of the unit. Therefore, each unit has 4 trainable coefficients $\mathbf{W}^{(l)}$ plus a trainable bias $b^{(l)}$, and all the units in a filter map share the same set of weights. A complete convolutional layer is composed of 64 filter maps, and each map uses different sets of weights and biases, thereby extracting different types of local features. These features are then combined by the subsequent layers in order to capture higher order features. For the last convolution, we take 16 filters to reduce the dimension of output, and the experimental results prove that it is superior to 64 filters.

With these convolutional operations, PCNN is able to extract the local temporal dependency among neighboring days and time slots and learn multi-grained features. Then we transmit these features $\mathbf{H}^{(L)}$ to the output layer to generate the predicted congestion level $\hat{c}_{m,n}$. Here we use the identity function as the activation function,
\begin{equation}
\hat{c}_{m,n} = \mathbf{W}^{o} \cdot \mathbf{H}^{(L)} + b^{o},
\end{equation}
where $\mathbf{W}^{o}$ is a weight term and $b^{o}$ is a bias term in the layer.

\subsection{Loss function}

We use the square error between the predicted congestion levels and the observed values to define the objective function, i.e.,
\begin{equation}
\label{loss}
\ell = \min \sum_{m=1}^{M} \sum_{n=1}^{N} \dfrac{1}{2} \|\hat{c}_{m,n} - c_{m,n}\|^{2} + \dfrac{1}{2} \lambda \|\mathbf{\Theta}\|^{2},
\end{equation}
where $M$ is the number of days in the training set and $N$ is the number of time slots in a day. $\mathbf{\Theta}$ represents the whole parameters in PCNN, and $\lambda$ is the regularization coefficient.

Note that, our proposed PCNN is able to make not only one-step ahead predictions, but also multi-step ahead predictions. Given a sequence of observed congestion data $\{ c_{i,j} \}$, $i = 0, 1, \ldots, m$, $j = 0, 1, \ldots, n-1$, when predicting $u$-step ahead congestion level $c_{m,n+u-1}$, we just take the $2t \times d$ values $(\langle c_{m-i,n+u-1-t}, \ldots, c_{m-i,n+u-2+t} \rangle, i \in \{ 1, 2, \ldots, d\})$ of the past $d$ days around the time slot $n+u-1$ and the $2t$ values $\langle c_{m,n-t}, \ldots, c_{m,n-1}, c_{m,n-1}, \ldots, c_{m,n-t} \rangle$ as the input of PCNN.
Then we compute $\hat{c}_{m,n+u-1}$ with PCNN and define the objective function according to Equation~(\ref{loss}).

\subsection{Algorithm and optimization}
We then use the stochastic gradient descent (SGD) method with the RMSprop update rule \cite{tieleman2012lecture} to minimize the square errors between our predictions and the actual congestion levels. Algorithm 1 outlines the training process of PCNN. We first construct the training instances from the original traffic congestion level data (lines 1-5), i.e., building the input matrix $\mathbf{X}_{m,n}$ for each predicted traffic congestion level $c_{m,n}$ in time slot $n$ of day $m$. Then, PCNN is trained via back propagation (lines 6-10).

\begin{algorithm}[h]
\small
\label{alg:pem}
\caption{PCNN Training Algorithm}
\begin{algorithmic}[1]
\REQUIRE historical congestion levels $\cal{C}$, the size of the input matrix, the number of convolutions;
\ENSURE the learned model;
\\// \textit{construct training instances}
\STATE $\cal{D} \leftarrow \emptyset$;
\FOR {$c_{m,n} \in \cal C$}
\STATE build the input matrix $\mathbf{X}_{m,n}$ according to Equation~(\ref{matrix});
\STATE put a training instance $(\mathbf{X}_{m,n}, c_{m,n})$ into $\cal{D}$;
\ENDFOR
\\// \textit{train the model}
\STATE initialize all parameters $\Theta$;
\REPEAT
\STATE randomly select a batch of instances ${\cal D}_{b}$ from $\cal{D}$;
\STATE update $\Theta$ by minimizing the objective (\ref{loss}) with ${\cal D}_{b}$;
\UNTIL {\textit{stopping criteria is met}}
\end{algorithmic}
\end{algorithm}

\section{Experiments}\label{experiment}
In this section, to evaluate the effectiveness of PCNN, we first introduce our dataset and basic settings, and then demonstrate the performances evaluated with different parameters. Finally, we show the experimental results compared with several baselines.

\subsection{Dataset and settings}

With the deployment of surveillance cameras on road networks, vehicles are photographed when they pass by, and structured vehicle passage records (VPRs) containing vehicle ID, location, and timestamp can be subsequently extracted from pictures using optical character recognition (OCR) \cite{chen2015mining}. The accuracy of recognizing the plate number by OCR could reach 97\% in ideal weather/lighting conditions. In our experiments, we collect six weeks of VPRs from 614 road segments in Jinan, China.

\begin{table}[!t]
\small
\caption{Definition of traffic conditions.}
\renewcommand{\arraystretch}{1.5}
\centering
\begin{tabular}{ l l }
\Xhline{1pt}
   group  &  description  \\
\hline
\textit{normal} traffic & $c \leq 1$   \\

\textit{light} congestion & $1 < c \leq 3$  \\

\textit{heavy} congestion & $c > 3$  \\
\Xhline{1pt}
\end{tabular}
\label{congestionLevel}
\end{table}

\subsubsection{Preprocessing}
In this study, we notice that traffic conditions on weekends clearly differ from those on workdays, and traffic jam rarely occurs on weekends (as indicated in Fig.~\ref{oneweek}); therefore we only use the traffic data on workdays (30 days in total) in our experiments. Further, as few people drive late at night, we only keep those records captured from 6:00 to 24:00 everyday. Here we first take 5 minutes as the size of time slot (later we compare different methods with various sizes of slot), and compute the traffic congestion levels for all the slots based on Definition 1. Thus, we have 3,978,720 ($614 \times 30 \times 18 \times (60/5)$) values of traffic congestion level. As forecast of congestion level is more important when traffic is heavy, we differentiate the traffic conditions based on the value of congestion level $c$, and define the traffic condition as \textit{congested} if $c$ is larger than 1, as shown in Table~\ref{congestionLevel}.

\begin{figure}[!t]
\centering
\subfigure[Cdf of congestion levels.]{
\includegraphics[width=0.22\textwidth]{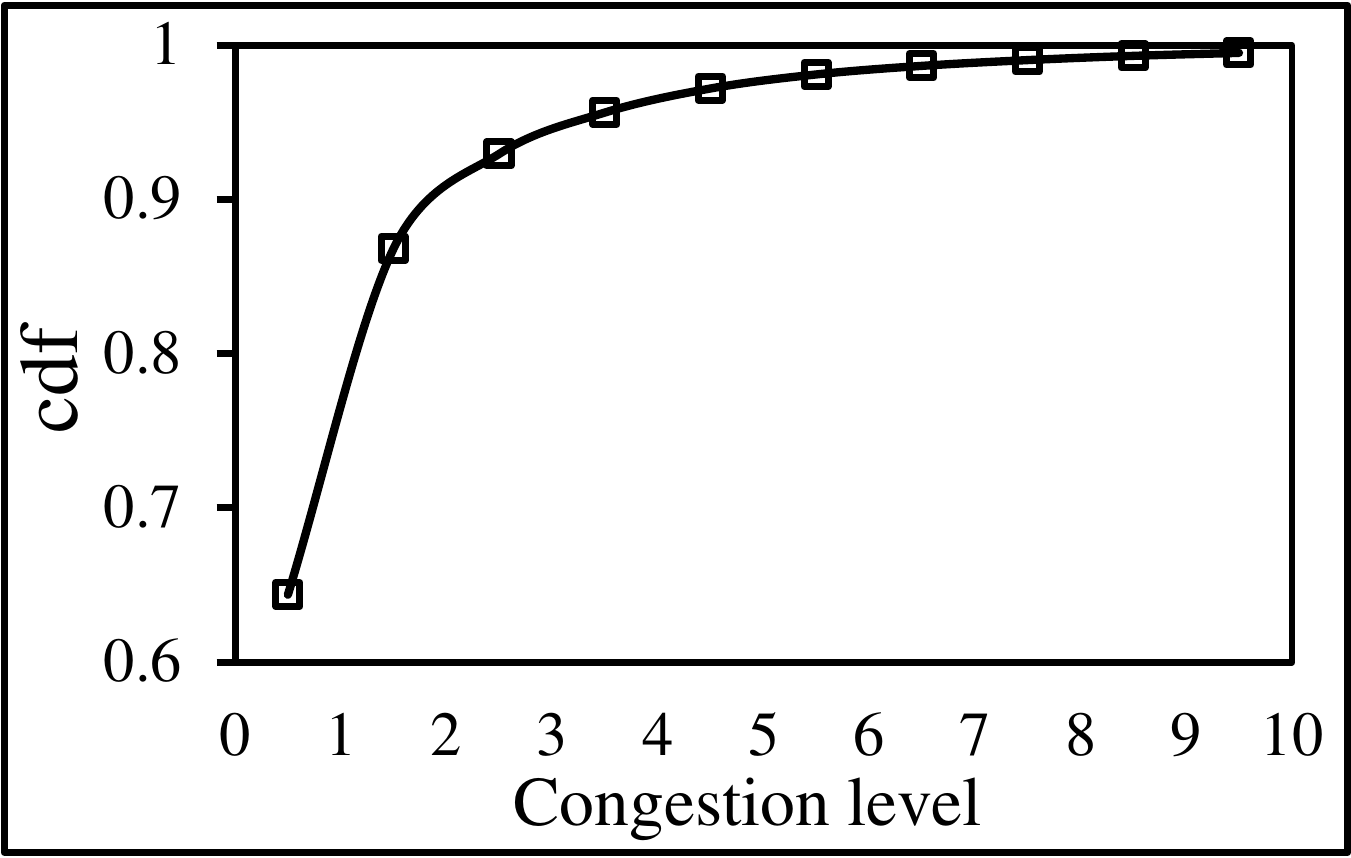}}
\subfigure[Distribution of \textit{congested} traffic.]{
\includegraphics[width=0.22\textwidth]{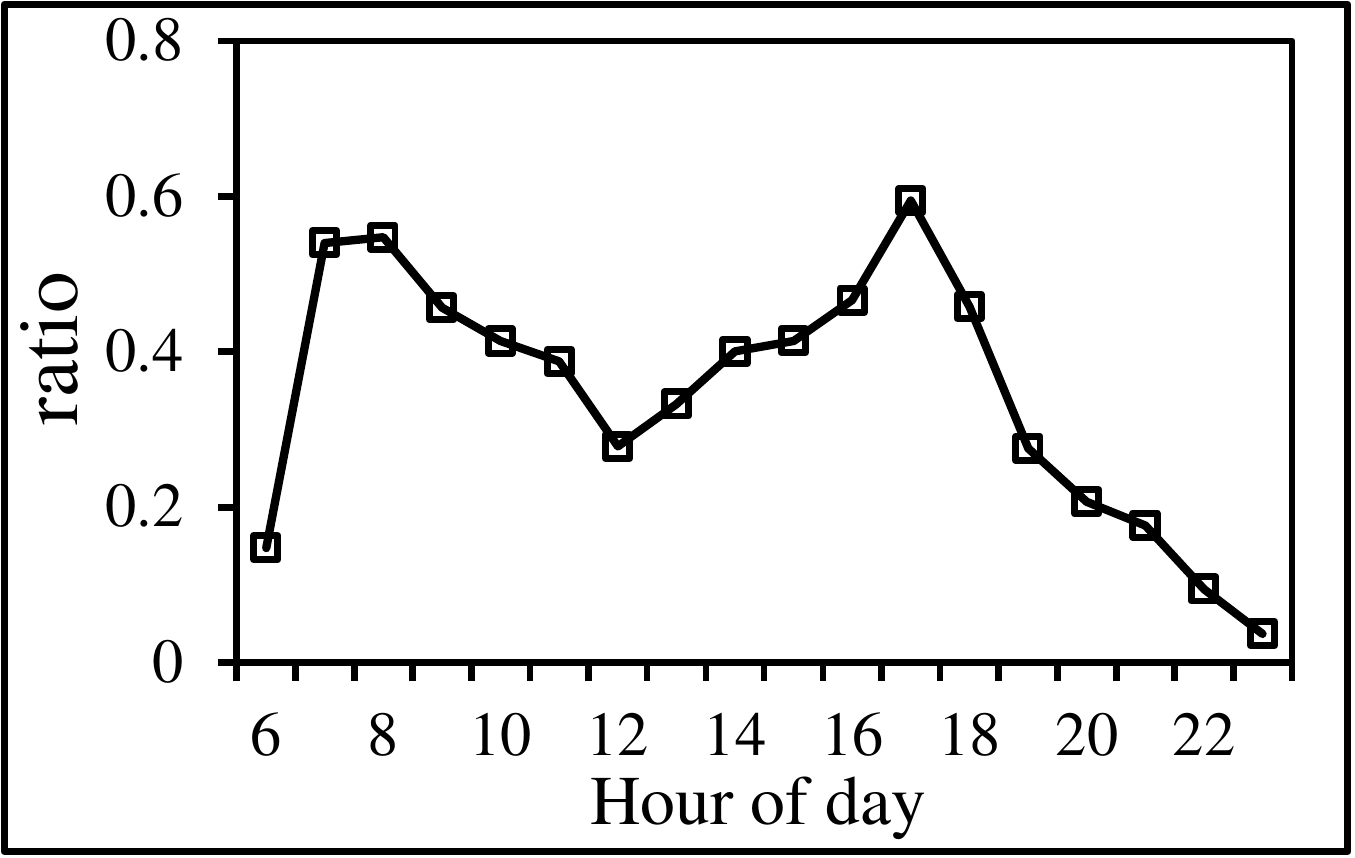}}
\caption{Characteristics of the traffic dataset.}
\label{statistic}
\end{figure}

\begin{figure*}[!t]
\centering
\subfigure[MAE ($d$=9)]{
\includegraphics[width=0.3\textwidth]{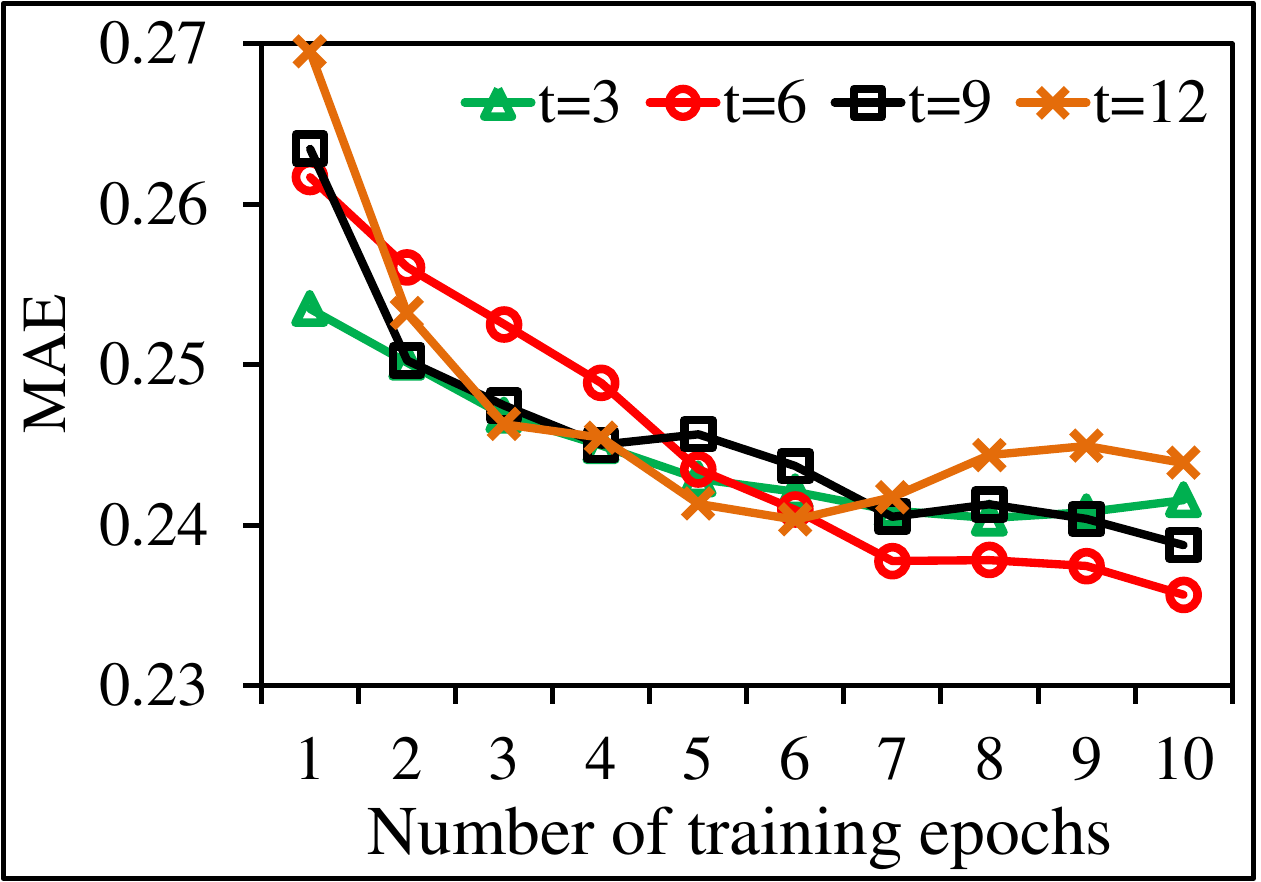}}
\subfigure[RMSE ($d$=9)]{
\includegraphics[width=0.3\textwidth]{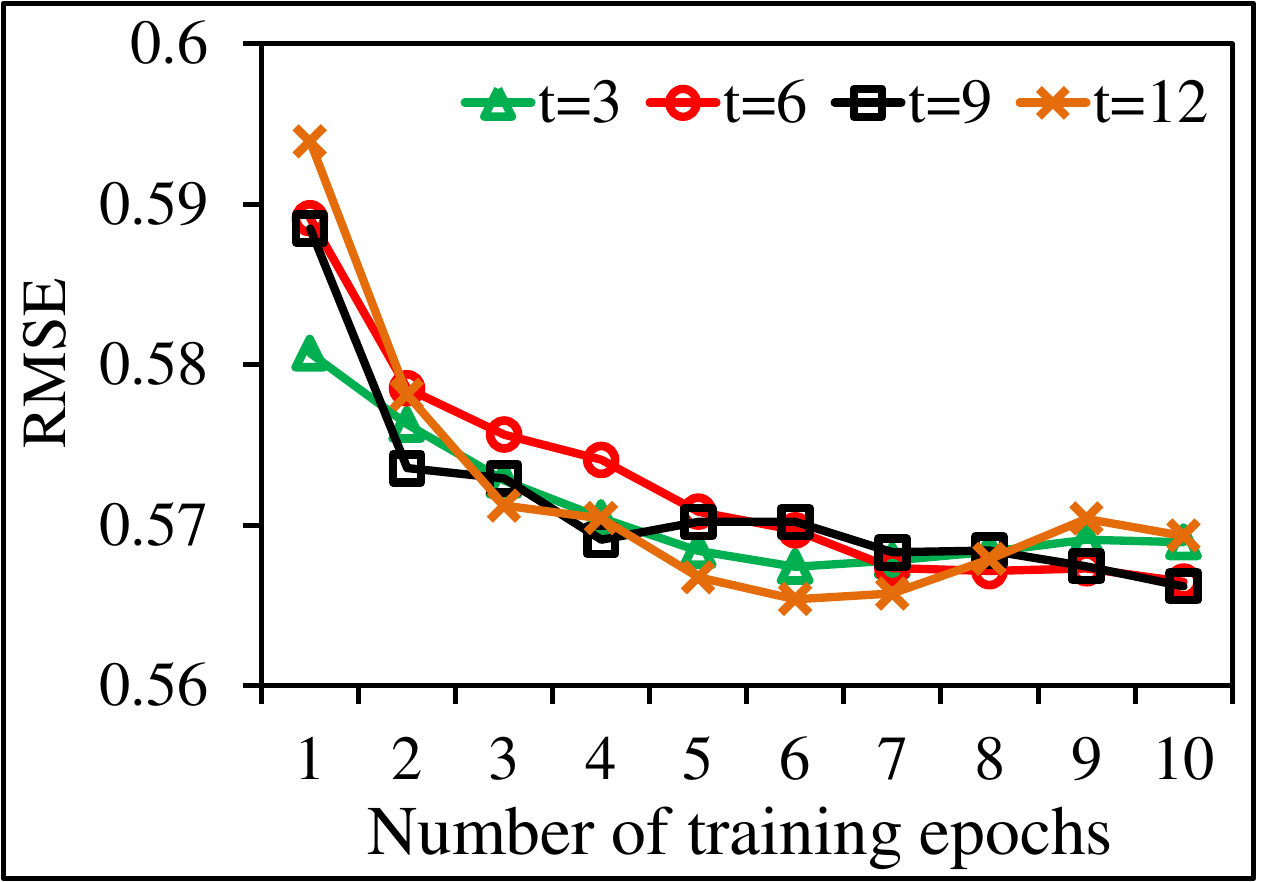}}
\subfigure[MRE ($d$=9)]{
\includegraphics[width=0.3\textwidth]{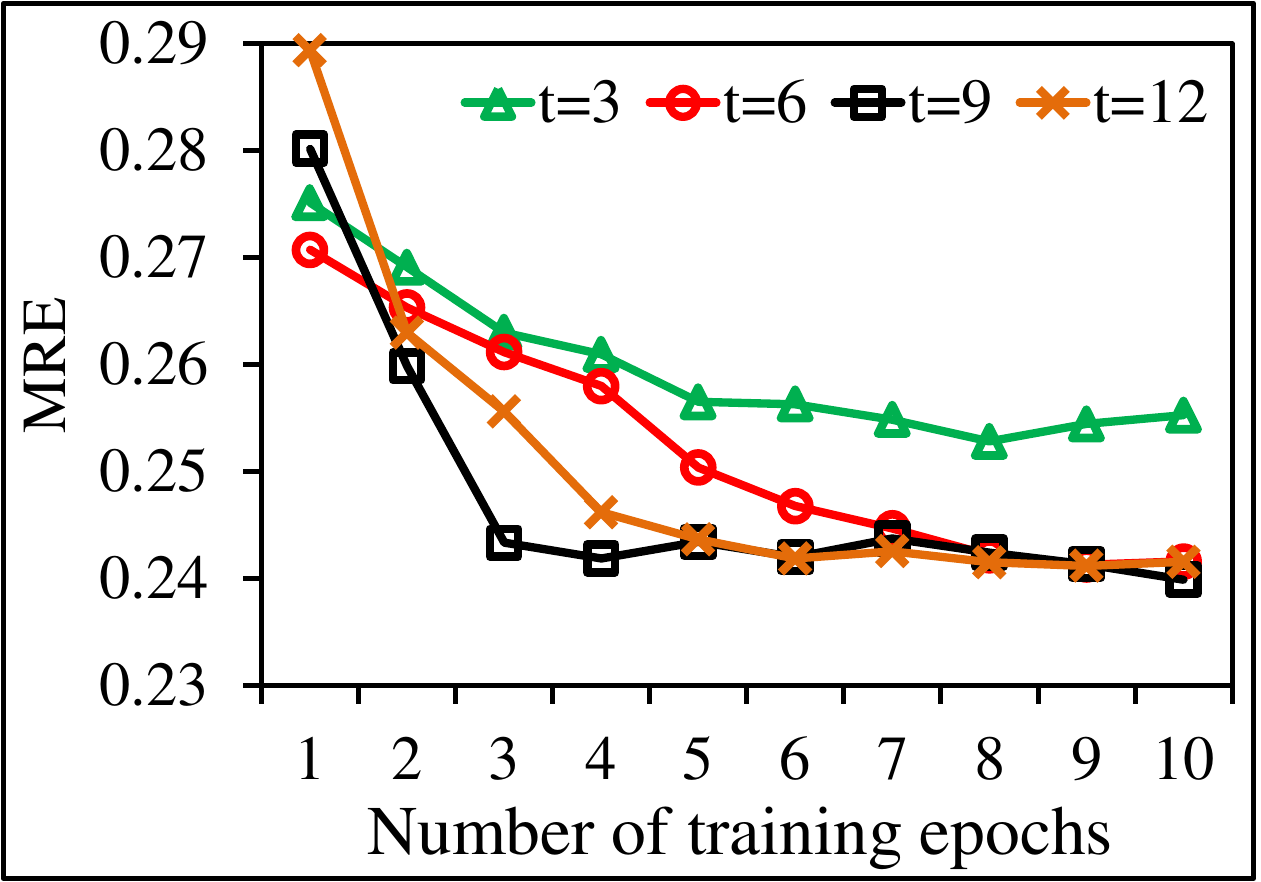}}
\subfigure[MAE ($t$=6)]{
\includegraphics[width=0.3\textwidth]{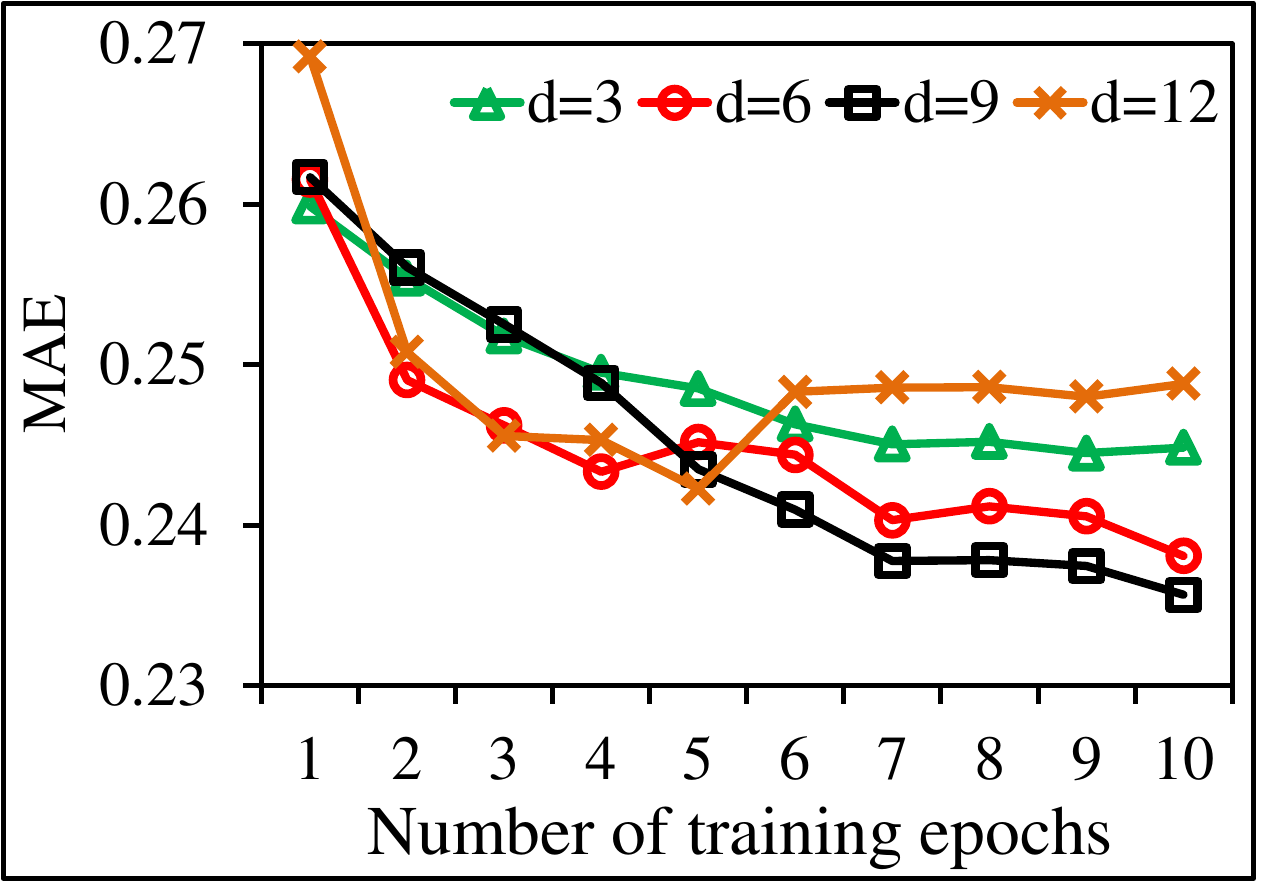}}
\subfigure[RMSE ($t$=6)]{
\includegraphics[width=0.3\textwidth]{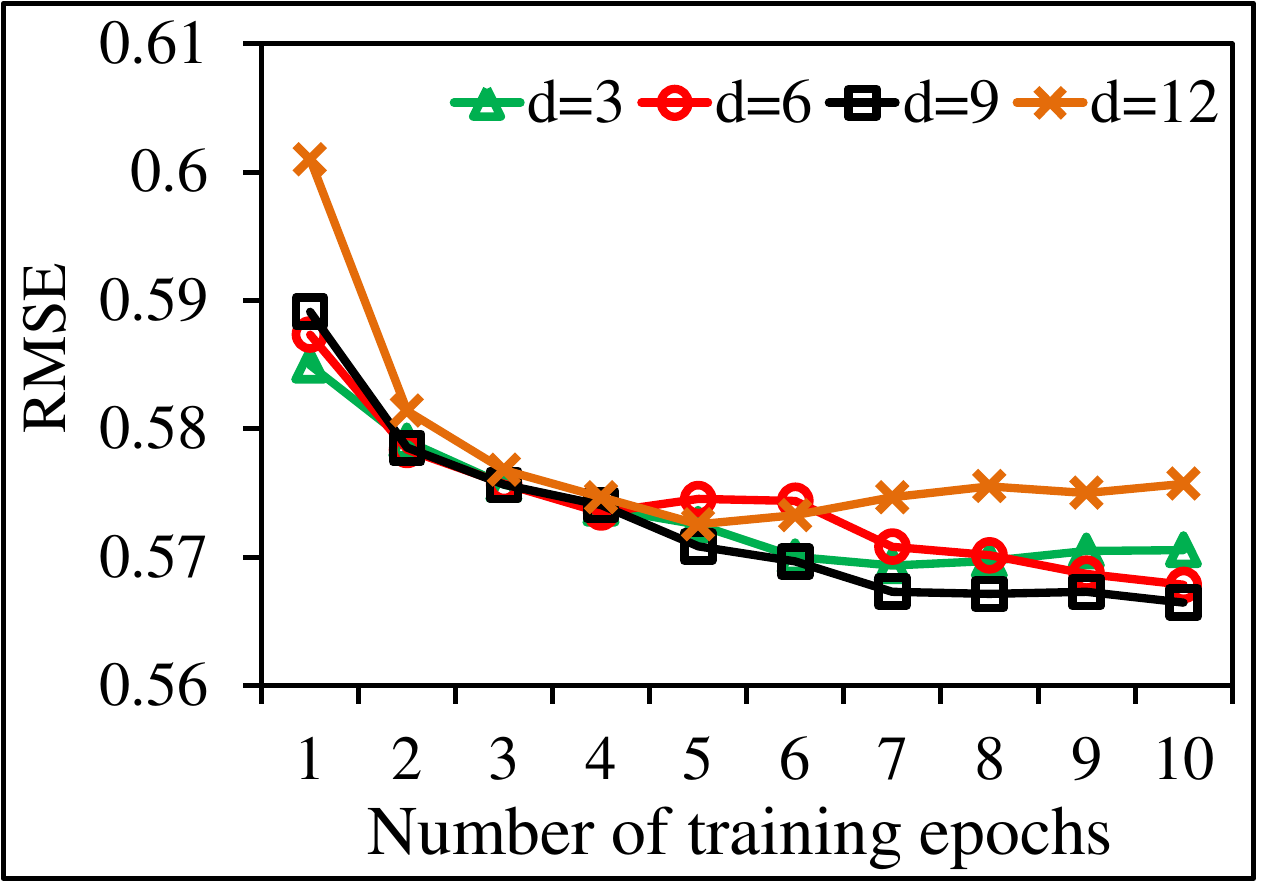}}
\subfigure[MRE ($t$=6)]{
\includegraphics[width=0.3\textwidth]{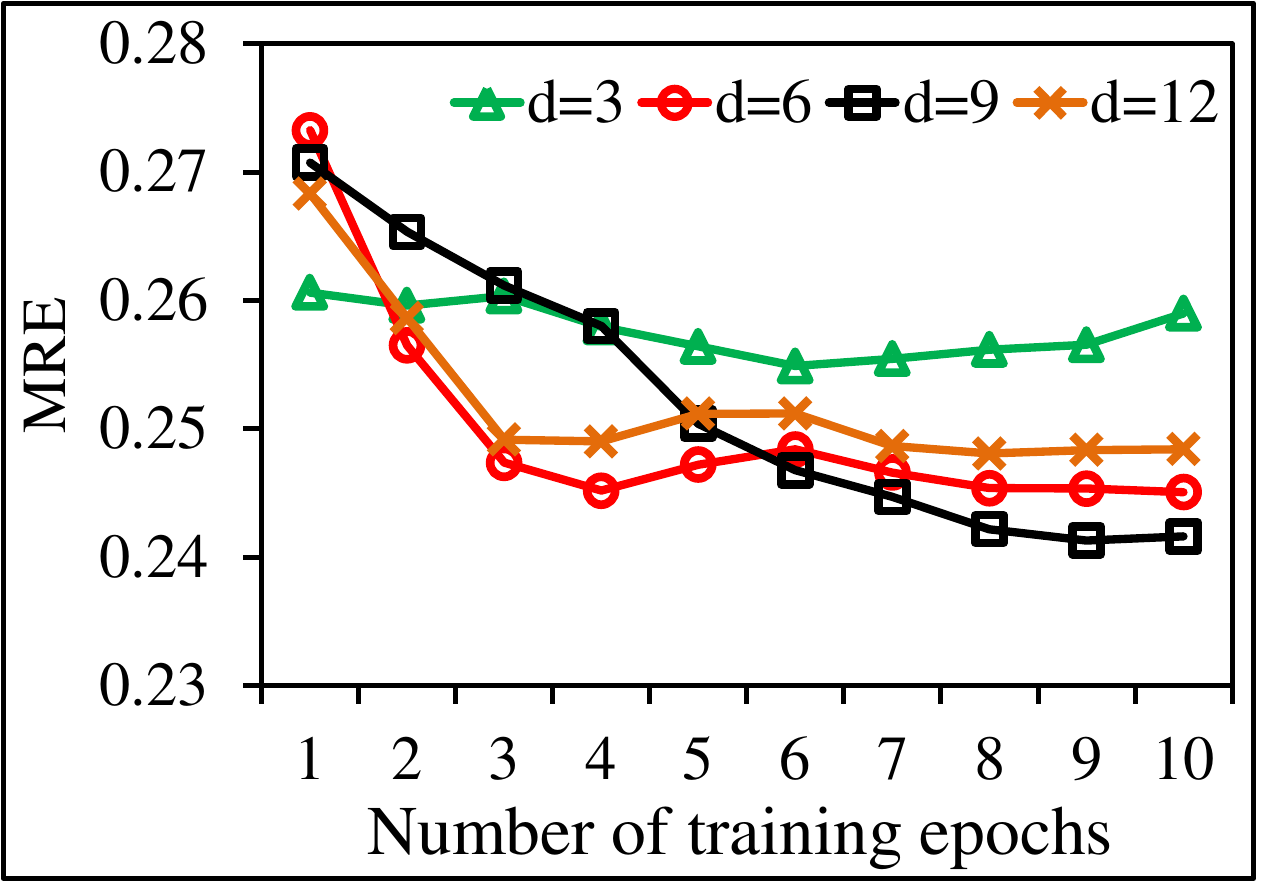}}
\caption{Effect of size of input matrix.}
\label{slot}
\end{figure*}

In order to understand the traffic congestion data better, we compute the cumulative distribution functions (cdf) of the congestion levels and the distribution of \textit{congested} traffic by time of day, as shown in Fig.~\ref{statistic}. Clearly, \textit{congested} traffic occurs in about 36\% time slots, and is mainly concentrated around the morning peak and the evening peak.

In the training process, we use the Min-Max normalization method to scale the whole dataset into the range $[0,1]$. In the evaluation, we re-scale the predicted values back to the normal values, to compare with the ground truth.

\subsubsection{Hyperparameters}
We use the open-source deep learning library, deeplearning4j \footnote{https://deeplearning4j.org/}, to build our models. The first $(L-1)$ convolutions use 64 filters of size $2 \times 2$, and the last one uses a convolution with 16 filters of size $2 \times 2$. We use ReLU as the activation function, and fix the learning rate at 0.005. We set the $l_{2}$ regularization parameter at 0.001, and the batch size at 128. These values are selected via a grid search on our dataset. We take the traffic congestion values of the first 20 days as the training set, and the next 5 days' data as the validation set for tuning parameters. Afterwards, we continue to train the model on the full dataset for a fixed number of epochs (e.g., 10 epochs), and compare the performance on the last 5 days' data with baselines.

\subsubsection{Evaluation metrics}
To evaluate the effectiveness of PCNN, we use three performance metrics, namely, the mean absolute error
(MAE), the root-mean-square error (RMSE), and the mean relative error (MRE), which are defined as
\begin{equation}
\tiny
\begin{aligned}
MAE &=  \dfrac{1}{M' \times N} \sum_{m=1}^{M'} \sum_{n=1}^{N} \| \hat{c}_{m,n} - c_{m,n} \|,\\
RMSE &=  \left[\dfrac{1}{M' \times N} \sum_{m=1}^{M'} \sum_{n=1}^{N} \| \hat{c}_{m,n} - c_{m,n} \|^{2} \right]^{\scriptscriptstyle{ \dfrac{1}{2}}},\\
MRE &= \dfrac{1}{M' \times N} \sum_{m=1}^{M'} \sum_{n=1}^{N} \dfrac{\| \hat{c}_{m,n} - c_{m,n} \|} {c_{m,n}},
\end{aligned}
\end{equation}
where $\hat{c}_{m,n}$ is the predicted congestion level, and $c_{m,n}$ is the observed value. $M'$ is the number of days in the test set and $N$ is the number of time slots in a single day.

\subsection{Performance of PCNN}

In this section, we first evaluate the performance of PCNN with different parameters, namely, the size of the input matrix (the number of days, $d$ and the number of time slots, $t$), and the number of convolutional layers ($L$), and tune them one by one on the validation set. Then we show the detailed forecast performance with respect to varying traffic conditions and time of day.

\begin{figure*}[!t]
\centering
\subfigure[MAE]{
\includegraphics[width=0.3\textwidth]{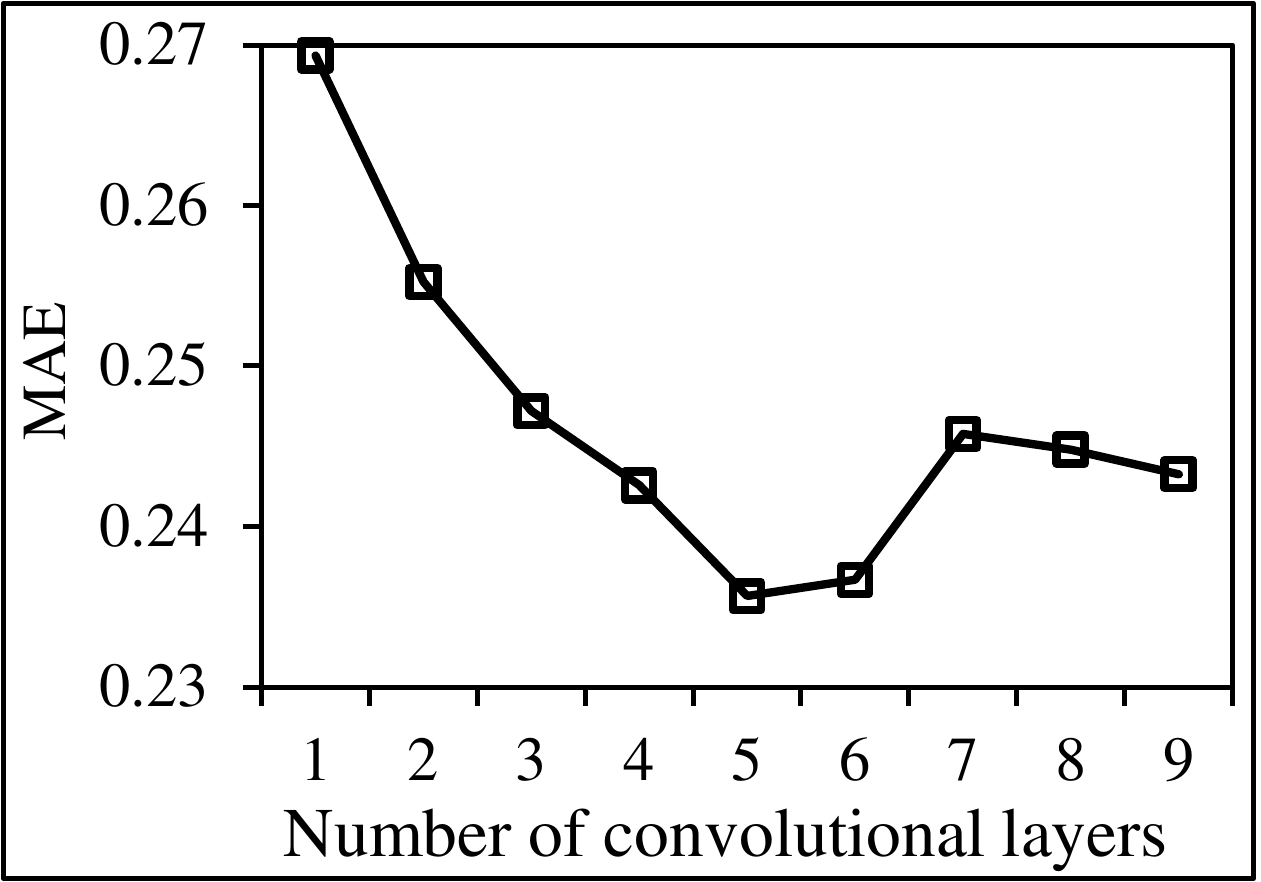}}
\subfigure[RMSE]{
\includegraphics[width=0.3\textwidth]{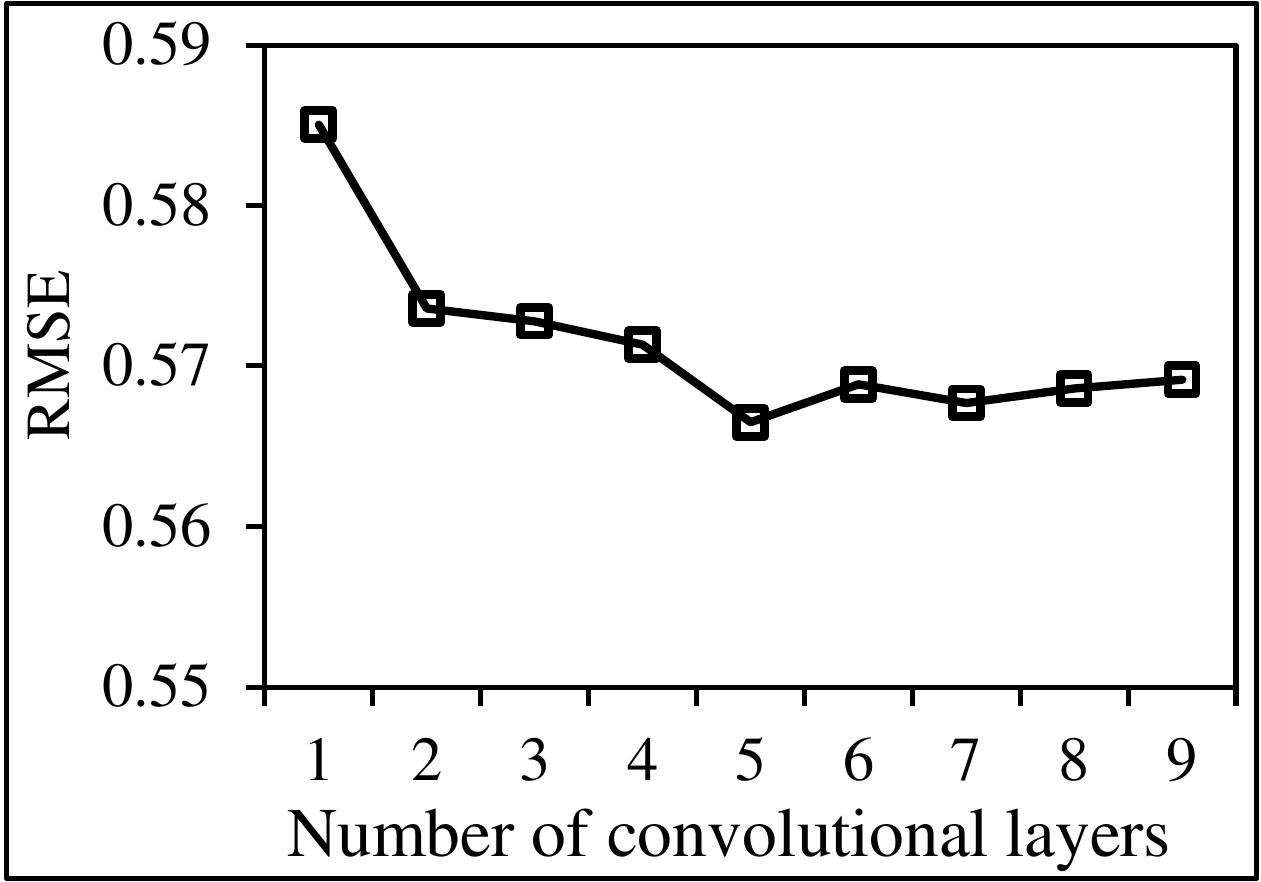}}
\subfigure[MRE]{
\includegraphics[width=0.3\textwidth]{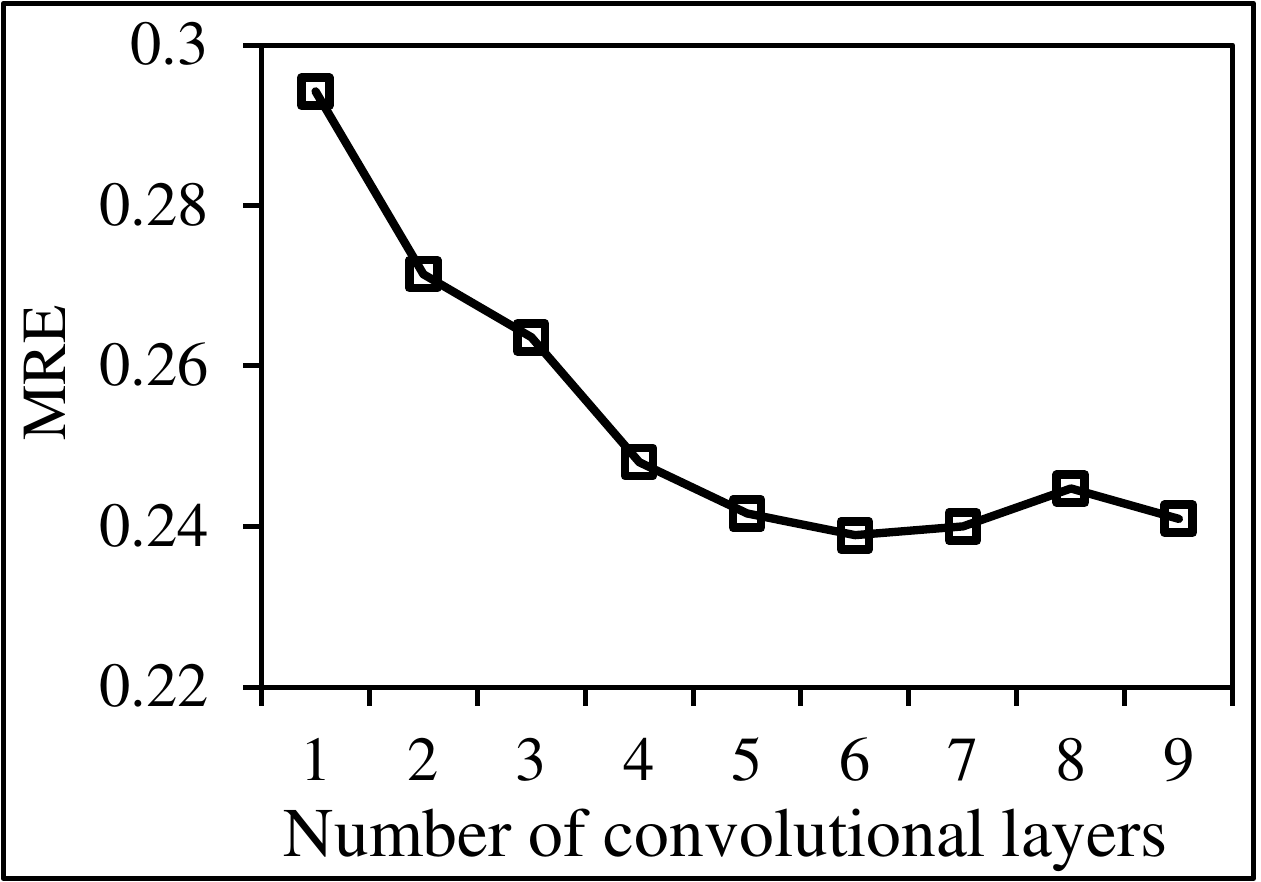}}
\caption{Effect of number of convolutional layers.}
\label{convolution}
\end{figure*}

\subsubsection{Identifying a suitable size of the input matrix}

On one hand, we know that the variations of congestion level in one day are similar to those in the preceding days. On the other hand, the congestion level is closely related to those in the adjacent slots. Thus, we set $t$ and $d$ at 3, 6, 9, 12 respectively in this case, and
choose the optimal number of convolutions for the models with different input matrix sizes. We first evaluate the effect of $t$ with defined $d$ on forecast accuracy, as shown in Fig.~\ref{slot} (a), (b) and (c). It can be observed that (1) with increase in the number of training epochs, the prediction errors (including MAE, RMSE, and MRE) start to decline, and remain stable after about 8 epochs; (2) the model with $t=3$ performs relatively poor, as it only considers the traffic conditions in the neighboring 15 minutes, without taking enough related values into consideration; (3) the model with $t=6$ performs the best, indicating that exploiting traffic conditions in half an hour around the current slot is the most suitable in our case. The impact of $t$ with other $d$ (i.e., $d=3,6,12$) on forecast accuracy is similar, and we set $t$ at 6 in the following experiments.

We then measure the impact of $d$ on forecast accuracy in terms of the three error criteria, as shown in Fig.~\ref{slot} (d), (e) and (f). Similarly, for different $d$, the prediction errors decrease when the number of epoches increases, and the model with $d=9$ obtains the best performance. In addition, the model involving the larger input matrix contains more parameters, and it needs more time to complete the training procedure. Therefore, we choose $t=6, d=9$ and 10 training epochs as our default setting, and the size of the input matrix is $10 \times 12$.

\subsubsection{Identifying the number of convolutional layers}
The number of convolutional layers determines the depth of PCNN, and we need to validate whether deep networks are more effective than the shallow ones. With the $10 \times 12$ input matrix, we consider a series of $L$ (the number of convolutions) values in this study, ranging from 1 to 9, and train the models with the same parameter setting. The experimental results are demonstrated in Fig.~\ref{convolution}. Clearly, a consistent improvement in forecast accuracy is observed with an increase in the number of convolutional layers, as the proposed model cannot capture enough multi-grained features with the shallow networks (e.g., 1 or 2 convolutions); on the other hand, the models with very deep networks (e.g., 8 and 9 convolutions) also get relatively large forecast errors, and take more time in the training process to train. Thus, we set the number of convolutional layers at 5 in our study based on the performance of prediction.

\subsubsection{Accuracy of forecast by different traffic conditions and time of day}
\begin{figure*}[!t]
\centering
\subfigure[MAE by level of traffic and hour of day]{
\includegraphics[width=0.7\textwidth]{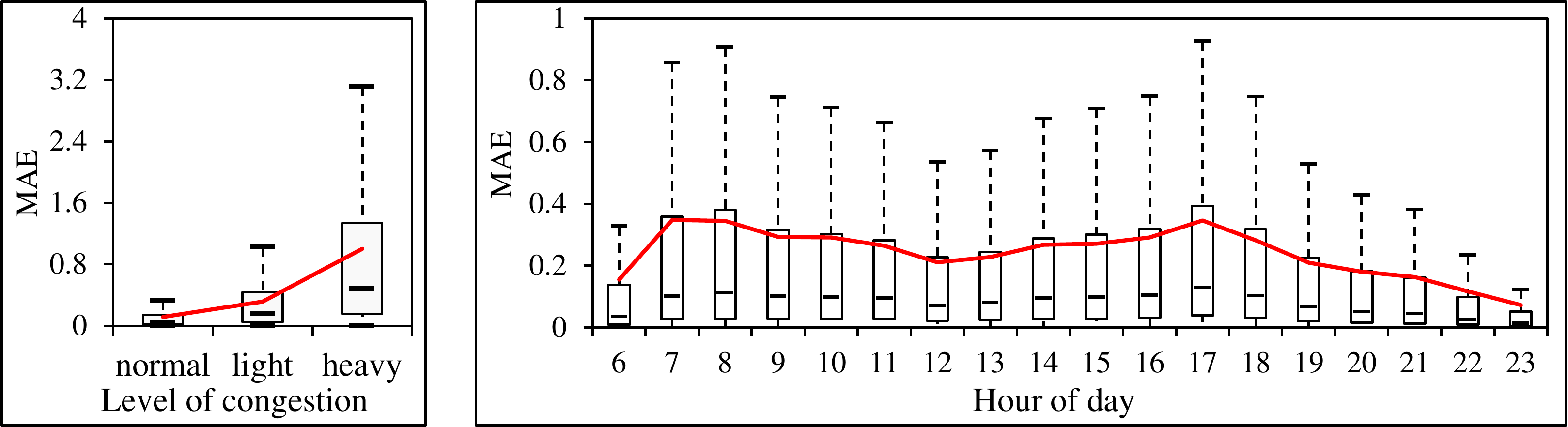}}
\subfigure[RMSE by level of traffic and hour of day]{
\includegraphics[width=0.7\textwidth]{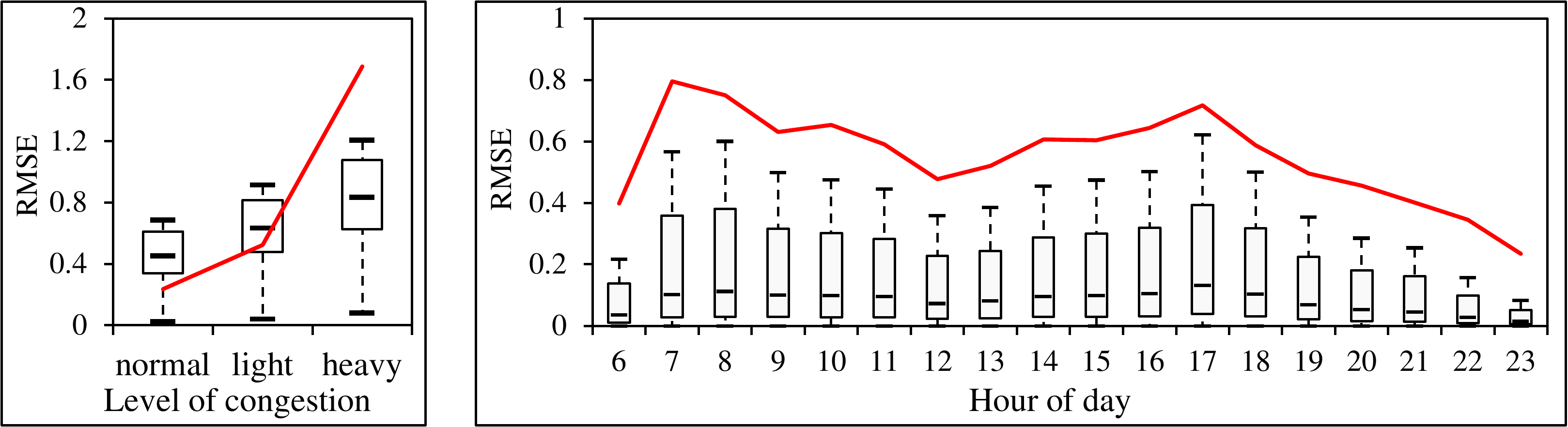}}
\subfigure[MRE by level of traffic and hour of day]{
\includegraphics[width=0.7\textwidth]{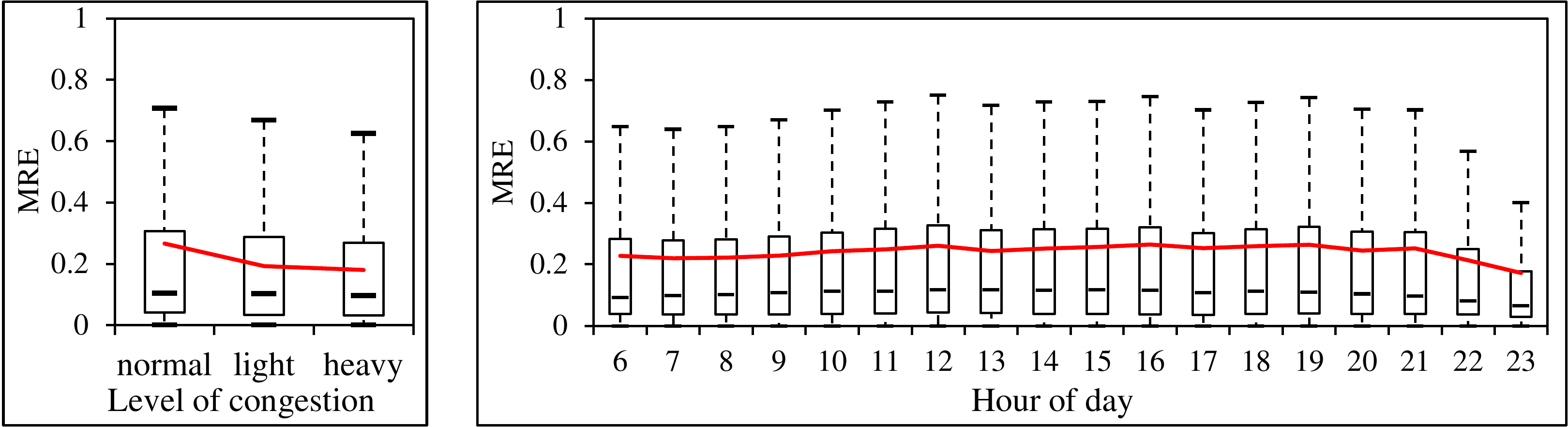}}
\caption{Forecast errors by level of traffic and time of day.}
\label{hour}
\end{figure*}

To evaluate the effectiveness of the proposed model further, we examine the performance by different traffic conditions and time of day. Fig.~\ref{hour} shows the performance of short-term congestion forecast using PCNN in terms of MAE, RMSE and MRE by various traffic conditions and hour of day. The box plots show the spread of the forecast errors and the red solid line represents the mean of the errors.

At forecast time, MAE and RMSE have consistent improvements when the traffic conditions change from normal to congested, as both MAE and RMSE consider only the magnitude of deviations of the forecasted values from the observed ones. Meanwhile, MRE provides a better sense of forecast accuracy as the errors are examined in terms of percentage deviations from the observed value, and it has consistent reduction evidently. Similarly, when the nature of the errors corresponding to the hour of day is examined, the forecast accuracy during peak-hours (7:00 - 9:00 and 17:00 - 18:00) is relatively lower when compared with off-peak hours. This is because of the fact that the patterns are more complicated and the values of congestion level are larger during peak-hours, as depicted in Fig.~\ref{statistic}.

\begin{figure}[!t]
\centering
\includegraphics[width=0.35\textwidth]{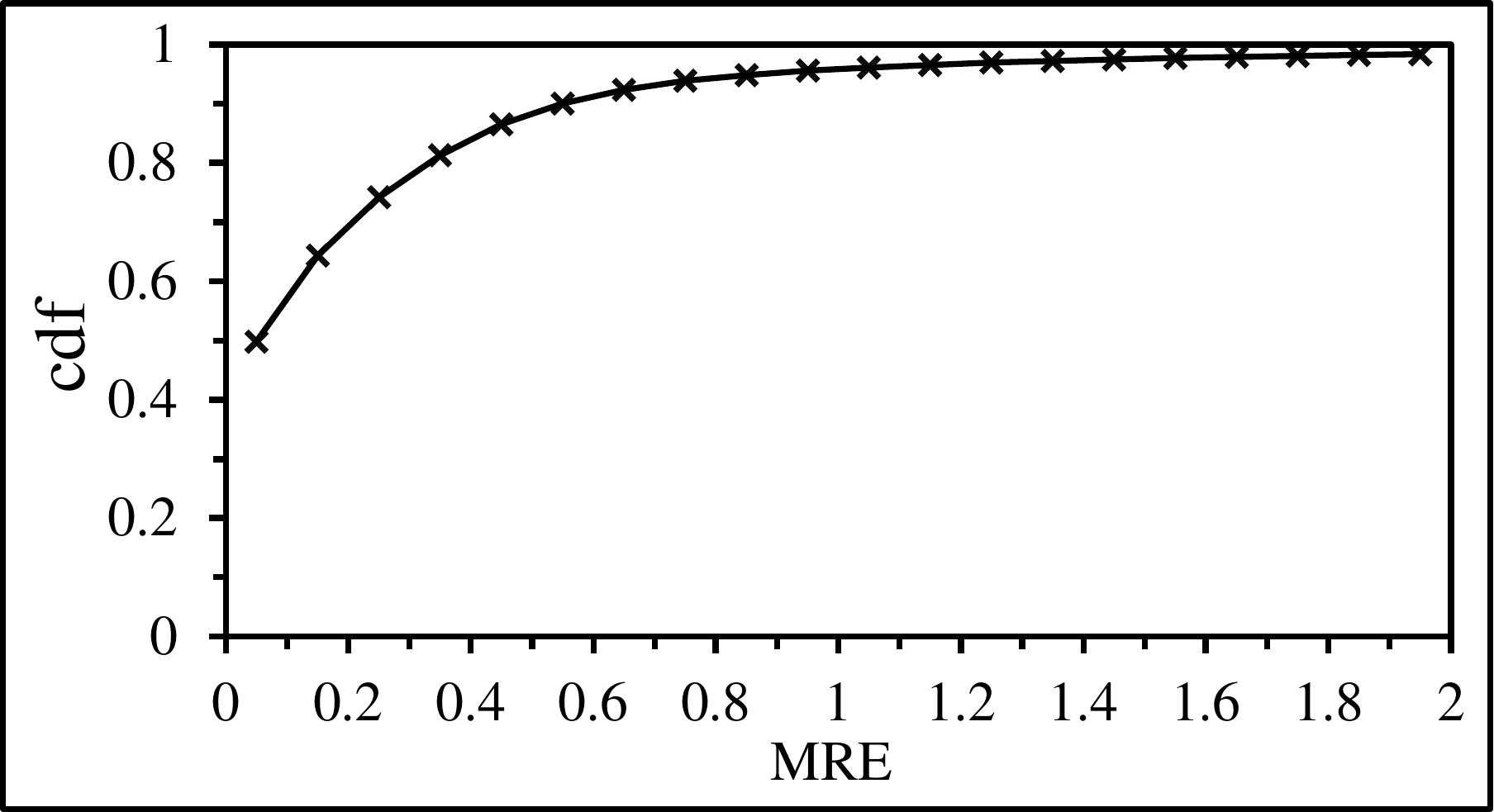}
\caption{Cdf of MRE.}
\label{mre}
\end{figure}

Moreover, examining the mean errors and the third quartiles (i.e., the top of the box), we find that the mean values are always greater than or approximately equal to the third quartiles, indicating that a few extremely large forecast errors exist in this case. To explore such cases, we investigate the distributions of errors, i.e., the cdf (cumulative distribution function) of MRE, as shown in Fig.~\ref{mre}. Clearly, about $74\%$ of the MREs are less than 0.3, and only $1.6\%$ of them are greater than 2. The reason may be that traffic accidents occur frequently, which often lead to a sudden surge in congestion levels within a short period of time. As instances with sudden change are rare, a general statistic model will be dominated by normal instances, and is difficult to capture the special patterns.

Generally speaking, according to the spread of MRE (the mean error is around 20\%) in Fig.~\ref{hour} (c), it can be said that the proposed PCNN provides reliable and accurate forecasts of traffic congestion levels.

\subsection{Comparisons with state-of-the-art methods}
To evaluate the performance of our proposal, we compare PCNN with several state-of-the-art methods for predicting short-term traffic congestion levels. To ensure a fair comparison, a common dataset and measure of performance are used.
\begin{itemize}
\item \textbf{HA:} This is probably the most straightforward method which assumes that the future value $\hat{c}_{m,n}$ is the average of the historical data.
\item \textbf{LR:} We use linear function on the input data to minimize the square error between our predictions and the actual values.
\item \textbf{ARIMA:} This is a general method for forecasting a time series, illustrated in detail in Section~\ref{relatedwork}.
\item \textbf{SARIMA:} We compare with the Seasonal ARIMA \cite{williams1998urban}, as traffic congestion levels have temporal periodicity.
\item \textbf{K-NN:} This is an enhanced K-nearest neighbors (K-NN) algorithm for short-term traffic forecasting based on identifying similar traffic patterns \cite{habtemichael2016short}.
\item \textbf{MLP:}  We construct many MLP structures with different numbers of hidden layers to predict traffic congestion values.
\item \textbf{LSTM:} This is a long short-term memory network which shows superior capability for time series prediction with long temporal dependency \cite{tian2015predicting}.
\end{itemize}


\begin{table*}[!t]
\centering
\caption{Results of methods.}
\label{tab:performance}
\renewcommand{\arraystretch}{1.2}
\small
\begin{tabular}{l |c c c |c c c |c c c |c c c}
\Xhline{1pt}
\multirow{2}*{method} & \multicolumn{3}{c|}{normal} & \multicolumn{3}{c|}{light} & \multicolumn{3}{c|}{heavy} & \multicolumn{3}{c}{overall}  \\

\multirow{2}*{} & MAE & RMSE & MRE & MAE & RMSE & MRE & MAE & RMSE & MRE & MAE & RMSE & MRE \\
\hline
HA(1) & 0.194 & 0.453 & 0.469 & 0.449 & 0.781 & 0.279 & 1.289 & 1.984 & 0.239 & 0.346 & 0.770 & 0.399 \\
				
LR(1) & 0.192 & 0.359 & 0.455 & 0.466 & 0.720 & 0.287 & 1.421 & 2.072 & 0.265 & 0.359 & 0.738 & 0.394 \\

ARIMA & 0.152 & 0.306 & 0.368 & 0.407 & 0.613 & 0.241 & 1.196 & 1.887 & 0.215 & 0.299 & 0.654 & 0.321\\

SARIMA & 0.149 & 0.298 & 0.361 & 0.392 & 0.601 & 0.233 & 1.180 & 1.854 & 0.212 & 0.293 & 0.636 & 0.313 \\

K-NN & 0.173 & 0.304 & 0.416 & 0.411 & 0.613 & 0.255 & 1.350 & 1.971 & 0.248 & 0.326 & 0.672 & 0.359 \\

MLP(1) & 0.147 & 0.276 & 0.340 & 0.369 & 0.571 & 0.226 & 1.141 & 1.730 & 0.210 & 0.282 & 0.600 & 0.299  \\

LSTM & 0.149 & 0.285 & 0.346 & 0.362 & 0.573 & 0.232 & 1.124 & 1.692 & 0.205 & 0.280 & 0.596 & 0.304  \\

\hline

HA(2) & 0.252 & 0.536 & 0.634 & 0.551 & 0.915 & 0.343 & 1.629 & 2.339 & 0.298 & 0.436 & 0.907 & 0.528 \\

LR(2) & 0.187 & 0.344 & 0.440 & 0.461 & 0.709 & 0.285 & 1.344 & 1.996 & 0.251 & 0.348 & 0.714 & 0.383  \\

MLP(2) & 0.139 & 0.259 & 0.321 & 0.351 & 0.533 & 0.217 & 1.102 & 1.676 & 0.197 & 0.268 & 0.573 & 0.283 \\

\hline
PCNN & \textbf{0.119} & \textbf{0.241} & \textbf{0.269} & \textbf{0.307} & \textbf{0.523} & \textbf{0.190}  & \textbf{0.947} & \textbf{1.636} & \textbf{0.172} & \textbf{0.232} & \textbf{0.557} & \textbf{0.240}  \\

\Xhline{1pt}
\end{tabular}
\end{table*}

\begin{table*}[!t]
\centering
\caption{Results of methods with different time granularity.}
\label{tab:granularity}
\renewcommand{\arraystretch}{1.2}
\small
\begin{tabular}{l |c c c |c c c |c c c |c c c}
\Xhline{1pt}
\multirow{2}*{method} & \multicolumn{3}{c|}{10-min} & \multicolumn{3}{c|}{15-min} & \multicolumn{3}{c|}{30-min} & \multicolumn{3}{c}{60-min}  \\

\multirow{2}*{} & MAE & RMSE & MRE & MAE & RMSE & MRE & MAE & RMSE & MRE & MAE & RMSE & MRE \\
\hline
LSTM & 0.263 & 0.596 & 0.268 & 0.259 & 0.601 & 0.273 & 0.254 & 0.592 & 0.264 & 0.253 & 0.586 & 0.259  \\

MLP(1) & 0.267 & 0.604 & 0.270 & 0.266 & 0.606 & 0.271 & 0.260 & 0.595 & 0.267 & 0.253 & 0.585 & 0.252  \\

MLP(2) & 0.256 & 0.571 & 0.261 & 0.252 & 0.567 & 0.257 & 0.244 & 0.562 & 0.252 & 0.231 & 0.557 & 0.243 \\

\hline
PCNN & \textbf{0.228} & \textbf{0.551} & \textbf{0.225} & \textbf{0.221} & \textbf{0.543} & \textbf{0.217}  & \textbf{0.214} & \textbf{0.541} & \textbf{0.212} & \textbf{0.210} & \textbf{0.536} & \textbf{0.207}  \\

\Xhline{1pt}
\end{tabular}
\end{table*}

To validate the necessity of folding the temporal data into a two-dimensional matrix, we experiment with both vectors and matrices as the inputs of HA, LR and MLP, denoted as HA(1), HA(2), LR(1), LR(2), MLP(1) and MLP(2). When predicting future congestion level $c_{m,n}$, the data in previous $t$ slots of day $m$ and in the $n$-th slot of preceding $d$ days are used to construct the one-dimensional vectors $\langle c_{m,n-t}, \ldots, c_{m,n-1}, c_{m-1,n}, \ldots, c_{m-d,n} \rangle $  of length $(t+d)$ as the inputs of HA(1), LR(1) and MLP(1). For the matrices, we flatten them and construct the new vectors $\langle c_{m-d,n-t}, \ldots, c_{m-d,n+t-1}, c_{m-d+1,n-t}, \ldots, c_{m,n-t} \rangle $  of length $(d+1)\times 2t$ as the inputs of HA(2), LR(2) and MLP(2).

For all the baselines, the system configuration parameters are optimized by a grid search. For instance, we vary $K$ from 5 to 50 at a step of 5 for K-NN, and find that it obtains the best performance when $K$ is 15; as it is very tricky to set up configuration options for MLP, we train a group of MLPs with the number of hidden layers varying from 1 to 9 and the number of neurons in each layer varying from 50 to 500, and the MLP structure with the best performance we can come up with contains 8 hidden layers with 150 neurons in each layer for the two-dimensional input and 5 layers with 200 neurons for the one-dimensional input.

To compare the proposed approach (PCNN) with the baselines, we not only evaluate the overall prediction performance, but also measure
it under different traffic conditions. We demonstrate the forecast errors on testing data in Table~\ref{tab:performance}, and our evaluation on the proposed method is 3-fold.
\begin{enumerate}
\item For one-dimensional input, we measure all baselines, including the naive method (HA), the regressive models (LR, ARIMA and SARIMA), the pattern recognition method (K-NN), and the neural networks (MLP and LSTM). Specifically, HA treats each element equally and LR assigns the elements with different weights, so LR performs better than the naive HA; ARIMA and SARIMA use the differencing step to eliminate the non-stationarity, and their forecast errors are smaller than LR's; SARIMA considers the temporal periodicity, and it performs better than ARIMA; different from the regressive models, K-NN identifies the similar patterns and predicts future value based on them, but it does not obtain decent performances, especially in the heavy congestion condition, as there is only 7\% of data in that case (as depicted in Fig.~\ref{statistic}); MLP uses many non-linear functions to model the relationship between the predicted values and the actual congestion levels, and performs better than K-NN; LSTM has an advantage of memorizing long historical data and can achieve lower prediction errors.
\item We then analyze the effect of introducing the two-dimensional input. HA averages more weakly related values when using the two-dimensional input, so it performs worse than with the one-dimensional input. LR and MLP with the two-dimensional perform better than those with the one-dimensional (according to the MRE metric, the error rates of LR(2) and MLP(2) decline by 2.8\% and 5.4\% respectively), as they could extract more effective features through complex linear or non-linear functions from the input matrix, validating the effectiveness of folding periodic time series data and constructing the two-dimensional input.
\item Our proposed method, PCNN, is clearly superior to the baselines based on the experimental results. Taking the overall prediction as an example, compared with the LSTM that has the best performance with one-dimensional input in the baselines, the MRE rate drops by 21.1\%, which is a considerable improvement; even compared with MLP(2) that has the same input matrix, the MRE of PCNN still drops by 15.2\%. The reason is two-fold: on one hand, the two-dimensional input matrix takes both the real-time traffic conditions and the historical similar traffic patterns into consideration; on the other hand, a series of convolutions over the input matrix could capture the local temporal dependency and model multiscale traffic congestion features.
\end{enumerate}

Further, we set the size of time slot at 10, 15, 30, and 60 minutes respectively, and compare the proposed PCNN with the baselines (as MLP and LSTM outperform other baselines obviously, we only compare PCNN with them). As shown in Table~\ref{tab:granularity}, as the size of time slot increases, the prediction errors decline. It is evident because the congestion levels become smoother when we consider larger time slot. Furthermore, the proposed PCNN performs better than these baselines under the circumstances of different time granularity, further validating the robustness of our model.

\section{Conclusion}\label{conclusion}
In this paper, we have proposed a novel method (\textbf{PCNN}) based on the convolution-based deep neural network modeling periodic traffic data to make short-term traffic congestion prediction. The accurate forecast could be used as a decision support tool for traffic operators to design an alternative traffic management strategy to avoid traffic jams. Considering the characteristics of urban traffic congestion data, we fold the time series data and construct a two-dimensional matrix. PCNN takes the matrix as its input, and models the local temporal dependency and multiscale traffic patterns through multiple convolutional operations. Finally, we evaluate the performances of PCNN on a real traffic dataset, and experimental results show that the proposed method outperforms state-of-the-art baselines significantly.


\bibliographystyle{IEEEtran}
\bibliography{wenxian}

\end{document}